\documentclass[floats,prd,aps,eqsecnum]{revtex4}
\usepackage{graphicx}
\usepackage{amsmath}
\usepackage{subfigure}
\newcommand{\be}{\begin{equation}}
\newcommand{\ee}{\end{equation}}
\newcommand{\bea}{\begin{eqnarray}}
\newcommand{\eea}{\end{eqnarray}}
\newcommand{\beaa}{\begin{eqnarray*}}
\newcommand{\eeaa}{\end{eqnarray*}}
\newcommand{\dis}{$\displaystyle}

\begin{document}
\title{Numerical Radiation Reaction for a Scalar Charge in Kerr Circular
Orbit}
\author{Samuel E. Gralla}
\affiliation{Yale University}
\author{John L. Friedman}
\affiliation{University of Wisconsin--Milwaukee}
\author{Alan G. Wiseman}
\affiliation{University of Wisconsin--Milwaukee}
\date{February 28, 2005}

\begin{abstract}
We numerically calculate the dissipative part of the self-force on
a scalar charge moving on a circular, geodesic, equatorial orbit
in Kerr spacetime.  The solution to the scalar field equation is
computed by separating variables and is expressed as a mode sum
over radial and angular modes.  The force is then computed in two
ways: a direct, instantaneous force calculation which uses the
half-retarded-minus-half-advanced field, and an indirect method
which uses the energy and angular momentum flux at the horizon and
at infinity to infer the force.  We are able to show numerically
and analytically that the force-per-mode is the same for both
methods. To enforce the boundary conditions (ingoing radiation at
the horizon and outgoing radiation at infinity for the retarded
solution) numerical solutions to the radial equation are matched
to asymptotic expansions for the fields at the boundaries.
Recursion relations for the coefficients in the  asymptotic
expansions are given in an appendix.
\end{abstract}

\maketitle

\section{Introduction: motivation and summary}
\label{sec:intro}

With sensitive ground-based gravitational wave detectors, such as
LIGO, GEO, TAMA and VIRGO \cite{ligo,geo,tama} in operation and
the launch of a space-based gravitational wave detector LISA
\cite{lisa} planned in the next decade, the need to accurately
model the waves emitted from inspiraling binary star systems has
become acute.

In the case of LIGO and the other ground-based detectors, the
sensitivity of the instrument is at a maximum around a few hundred
Hertz. This frequency is comparable to the orbital frequency of
two neutron stars just prior to their final coalescence, and
therefore LIGO is particularly sensitive to these sources. These
constituents of these systems have comparable mass and relatively
weak gravitational fields; therefore the emitted signal can be
computed using a weak-field, slow-motion approximation (i.e a
post-Newtonian approximation \cite{biww}). Although this method --
and other improvements to the post-Newtonian method \cite{bcv} --
give waveforms that are accurate enough to detect signals buried
in the LIGO noise, a more detailed knowledge of late inspiral and
merger, accessible only by numerical evolution, is needed to
extract from the waves the astrophysics of their sources.

LISA, however, will be sensitive to signals with much lower
frequency, in particular, to waves from a stellar-size black hole
spiraling in to a (super-)massive black hole [$10^3-10^6M_\odot$].
Detectable sources with this extreme mass-ratio will have periods
of (many) minutes and will persist for weeks (or even months). In
addition, the smaller mass may spend part of its orbit deep in the
strong gravitational field of the larger mass, making a
post-Newtonian approximation inappropriate for computing the
predicted waveforms.  Clearly, this problem is better suited to
black hole perturbation theory where the background geometry
generated by the larger mass $M$ is treated exactly, and the
smaller mass $\mu$ generates a small perturbation of the geometry.
To lowest nontrivial order in $\mu/M$, the field generated by the
smaller mass causes its trajectory to deviate from a geodesic of
the background spacetime. This is the origin of the self-force.
Heuristically, one can think of the smaller mass as traveling on a
geodesic of the perturbed spacetime.  Although this description of
the force is intuitive and compelling, computing the self-force
entails  a number of conceptual and technical difficulties.

At the center of the conceptual difficulties is the {\it
renormalization} problem: even though the perturbing mass is small
compared to the central mass, the perturbed field diverges at the
position of the particle.  A number of authors
\cite{Dirac38,DB60,Hobbs68,MST97,QW97,Quinn00} have addressed this
difficulty and developed formal methods for removing the
divergence while preserving the finite parts of the field that
give rise to the self-force.  These papers solve the problem in
principle; there remain, however, a number of difficulties in
implementing these prescriptions to find the force and subsequent
motion and waveform.  In particular, there remains a {\it gauge}
problem. The self-force equations for a small mass moving in a
background spacetime have only been written down in Lorentz gauge,
but, unfortunately, most methods for finding the perturbing
gravitational fields use a different gauge. Complicating
matters further, recent calculations \cite{BO01} suggest that
the gauge transformations that relate gauges of perturbation
theory to Lorentz gauge are poorly behaved.  Thus it is difficult
to use the metric perturbations that are readily available from
black hole perturbation theory in the formal equations for the
self-force.

One method of side-stepping both the gauge and the renormalization
issue is to compute the self-force indirectly by computing the energy
and angular momentum flux at infinity.  Such techniques usually assume
that the particle is in circular geodesic orbit. The particle is then
the source of the metric perturbation (or of the electromagnetic or
scalar field perturbation).  The perturbed field is examined at the
horizon and infinity and the energy and angular momentum  flux is
computed. The rate of energy loss is then equated to an orbital energy
loss and the rate of inspiral can be inferred.  Although such
techniques have been broadly applied (e.g.
\cite{hughes01,poissonSasaki95}) they have a drawback: They give only
the time-averaged dissipative force. [See \cite{wiseman00} for a
discussion of the shortcomings of the time-averaged force.] When used
to infer the rate of orbital decay, these energy balance arguments
assume that the field energy on  a spatial slice is constant as we go
from one orbit to the other. See Section IVA for discussion.

In this paper, we attack a somewhat simpler problem that still
puts us on a path to solve the more general problem.  We assume
that our particle is a scalar charge that acts as the source of a
Klein-Gordon field. The self-force arising from the back-reaction
of a scalar (rather than a gravitational) field involves no
delicate gauge issues. The part of our work reported here avoids
the renormalization problem by using the
half-retarded-minus-half-advanced field.  The divergent structure
of the field at the position of the particle is the same for the
advanced and retarded solutions, and therefore the difference is
smooth at the particle's position. By evaluating the gradient of
the half-retarded-minus-half-advanced field we are able to compute
the dissipative (time-antisymmetric) parts of the self-force
acting on the particle at any instant; our method does not require
time-averaging to compute the force.  Although in the present
paper we will assume that our source particle is traveling on a
circular geodesic orbit, it is not an essential feature of the
method. Extending the method to a particle in non-circular orbit
would severely complicate the numerical calculation of the field,
but it would not appreciably complicate the calculation of the
self-force. However, in this special case of circular motion, we
are able to show that our direct calculation of the instantaneous
force agrees -- mode-by-mode -- with the indirect calculation
based on the average energy loss at infinity and at the horizon.

  The next steps in this work are to perform calculations for
non-circular orbits and to calculate the conservative as well as
the dissipative part of the self-force. As discussed above, the
only complication arising in the former task is the numerical
calculation of the field from an orbit that has a countably
infinite set of frequencies.  The latter, however, requires a new
procedure that will include a renormalization of the divergent
field.  It is in anticipation of the delicate subtraction arising
in mode-by-mode renormalization that we have taken such care with
the accuracy of the radial functions computed in this paper.

{\sl Outline and conventions}

In section \ref{Solution of the Wave Equation}, we use separation
of variables to solve the wave equation and write the scalar field
$\Phi$ as a sum over radial and angular mode functions. In section
\ref{Integration of the Radial Equation}, we describe the scheme
for the numerical construction of the radial functions which
appear in this solution.  In section \ref{Radiation Reaction}, we
present the argument that the two methods are equivalent, and
derive separate expressions for the evolution of the conserved
quantities $E$ and $L$ based on each one.  Finally in section
\ref{Discussion} we discuss numerical details, show the properties
of the radial functions, and present our results for dissipative
self-force.

Throughout the paper we use Boyer-Lindquist coordinates
$(t,r,\theta,\phi)$ and a metric with signature $(- + + +)$. The
mass of the black hole is $M$, its spin is $a$, and the usual
abbreviations $\Delta = r^2 + a^2 -2 M r$ and $\Sigma = r^2 +a^2
\cos \theta$ are used; the radial coordinate of the horizon is
written $r_+ = M + \sqrt{M^2-a^2}$. Our conventions are those of
Misner, Thorne and Wheeler \cite{mtw}, except that their symbol
for $\Sigma$ is $\rho^2$.


\section{Solution of the Wave Equation}
\label{Solution of the Wave Equation}

We consider the Klein-Gordon field $\Phi$ of a point paticle of
conserved scalar charge $q$, orbiting on a circular, equatorial
geodesic of the Kerr geometry. The field $\Phi$ satisfies the
massless scalar wave equation
\begin{equation}\label{Scalar Wave Equation}
\nabla _\alpha  \nabla ^\alpha  \Phi  =  - 4\pi \rho,
\end{equation}
with the scalar charge density $\rho$ a delta function along the
trajectory $z^\alpha(\tau)$.
\begin{equation}\label{Point Particle Source}
\rho  = q\int {\delta ^4 \left( {x^\alpha , z^\alpha  (\tau )}
\right)d\tau }.
\end{equation}
With this normalization, we have $\displaystyle q =\int_V \rho\,
dV$, where $dV$ is the volume element on a spacelike hypersurface
$V$ orthogonal to the trajectory. For a circular orbit of radius
$r_0$ and frequency $\Omega$, we have
\begin{equation}\label{Point Particle Source--Coordinates}
\rho  = q\int {\frac{{\delta \left( {t - u^t \tau } \right)\delta
\left( {r - r_0 } \right)\delta \left( {\theta  - \frac\pi2}
\right)\delta \left( {\phi  - \Omega u^t \tau} \right)}}{{\Sigma
\sin \theta }}d \tau},
\end{equation}
where $u^t$ is the (constant) time component of the particle
four-velocity, and Eq.(\ref{Scalar Wave Equation}) takes the form,
\begin{equation}\label{SCW With Source}
\nabla _\alpha  \nabla ^\alpha  \Phi  =  - 4\pi \frac{q}{{r_0 ^2
u^t }}\delta (r - r_0 )\delta (\theta  - {\textstyle{\pi  \over
2}})\delta (\phi  - \Omega t).
\end{equation}
Separating variables yields the solution
\begin{equation}\label{Field Sum}
\Phi  = \frac{1}{{2\pi }}\frac{q}{{u^t }}\sum\limits_{l,m} {R_{lm}
} (r)S_{lm} ({\textstyle\frac\pi2})S_{lm} (\theta )e^{im(\phi  -
\Omega t)},
\end{equation}
where the $S_{lm}$ and $R_{lm}$ satisfy angular and radial
equations given below. The $S_{lm}$ are oblate spheroidal
harmonics, given by

\begin{equation}\label{Angular Equation}
\left[ {\frac{1}{{\sin \theta }}\frac{\partial }{{\partial \theta
}}\left( {\sin \theta \frac{\partial }{{\partial \theta }}}
\right) - \frac{{m^2 }}{{\sin ^2 \theta }} + \lambda _{lm}  + m^2
\Omega ^2 a^2 \cos ^2 \theta } \right]S_{lm} \left( \theta
\right) = 0.
\end{equation}

We take the harmonics to be real, and normalize by
\begin{equation}\label{Spheroidal Normalization}
\int_0^\pi {S_{lm} (\theta )^2 \sin (\theta )d\theta }  = 1.
\end{equation}
We fix the sign by demanding that $S_{lm}e^{im\phi}=\sqrt {2\pi
}Y_{lm}$ when $a=0$ ($\lambda_{lm}$ will reduce to $l(l+1)$ in
this case). The numerical calculation of the $S_{lm}$ and their
eigenvalues $\lambda_{lm}$ is a straightforward task, and is not
discussed here. (See Numerical Recipes \cite{numrec} for a
treatment similar to ours.) The computation of the $R_{lm}$ is
considerably more complicated and is detailed below.

\section{Integration of the Radial Equation}\label{Integration of
the Radial Equation}

The radial functions satisfy
\begin{equation}
\left[ {\frac{\partial }{{\partial r}}\left( {\Delta
\frac{\partial }{{\partial r}}} \right) + \frac{{m^2 }}{\Delta
}\left( {\Omega ^2 \left( {r^2  + a^2 } \right)^2  - 4\Omega Mar +
a^2 } \right) - m^2 \Omega ^2 a^2  - \lambda _{lm} }
\right]R_{lm}\left( r \right) = -4 \pi \delta \left( {r - r_0 }
\right). \label{Radial Equation}\end{equation}
First found by Carter \cite{carter68}, this is also the Teukolsky
equation with spin $s=0$, specialized to our source (this equation
can be derived from Teukolsky (\cite{TeukolskyPaper}) by setting
$\omega = m \Omega$). To provide accurate initial data for
numerical integration of this equation, we solve the equation in
asymptotic series valid near the horizon and near infinity.  In
terms of the angular velocity of the horizon, $\displaystyle
\omega_+ = {a \over {2Mr_+}}$, and the ``tortoise coordinate"
$r$*, satisfying
\begin{equation}\label{rstar}
\frac{{dr^*}}{{dr}} = \frac{{r^2  + a^2 }}{\Delta },
\end{equation}
the series take the form
\begin{align}\label{Radial Series Horizon}
R_{lm}^ +   &= \sum\limits_{n = 0} C_n^ +  \left( {r - r_ +  }
\right)^n e^{ - im(\Omega  - \omega _ +  )r^*}\qquad
    &(\mbox{near } r_+),\\
\label{Radial Series Infinity} R_{lm}^\infty   &= \sum\limits_{n =
1} \frac{C_n^\infty}{r^n} e^{ im\Omega r^*} \qquad &(\mbox{near }
\infty).
\end{align}

   The signs in the exponentials of these series amount to a choice of
boundary conditions. Here we have picked ingoing radiation ($-$)
at the horizon and outgoing radiation (+) at infinity, to
construct the retarded solution.  Note that in the case of
circular orbits, where the time dependence is just $e^{-i m \Omega
t}$, observers at infinity and the horizon always agree on the
direction of the radiation down the hole, and ($-$) is always the
correct sign for this solution.

  The coefficients $C_n^+$ and $C_n^\infty$ satisfy recursion
relations given in the appendix, with overall normalization set
for each $l,m$ harmonic by the $\delta$-function source. The
asymptotic series provide initial values of $R_{lm}$ and its
derivative for numerical integration of the homogeneous radial
equation out from the horizon and in from infinity. We thereby
obtain homogeneous solutions $R^+_{lm}$ (from the horizon to
$r_0$) and $R^\infty_{lm}$ (from $r_0$ to infinity).  To obtain
the solution $R_{lm}$ to the inhomogeneous wave equation
(\ref{Radial Equation}), we patch $R^+_{lm}$ and $R^\infty_{lm}$
at $r=r_0$.  Requiring that $R_{lm}$ be continuous and that the
discontinuity in $dR_{lm}/dr$ be fixed by the $\delta$-function
source, we have
\begin{equation}\label{Continuity}
R_{lm}  = C_{lm} R_{lm}^ +  (r_ <  )R_{lm}^\infty  (r_ >  ),
\end{equation}
where $r_<$ ($r_>$) denotes the lesser (greater) of $r_0$ and $r$,
and
\begin{equation}\label{Patch}
C_{lm}  = \frac{-4 \pi}{{\Delta \cdot W(R_{lm}^ + ,R_{lm}^\infty
)}}.
\end{equation}
It is easy to show that the quantity in the denominator, $\Delta$
times the Wronskian of the integrated solutions, is a constant for
any two solutions of the homogeneous radial equation.  It will be
convenient to define constant factors $C_{lm}^+$ and
$C_{lm}^\infty$ for which $R_{lm} = C_{lm}^+R_{lm}^+$ when $r<r_0$
and $R_{lm} = C_{lm}^\infty R_{lm}^\infty$ for $r>r_0$, namely

\begin{align}\label{Clm Definition 1}
C_{lm}^+ &= \frac{{-4 \pi R^\infty_{lm}(r_0) }}{{\Delta W(R_{lm}^
+ ,R_{lm}^\infty  )}}\\
\label{Clm Definition 2} C_{lm}^\infty &= \frac{{-4 \pi
R^+_{lm}(r_0) }}{{\Delta W(R_{lm}^ + ,R_{lm}^\infty  )}}.
\end{align}
\section{Radiation Reaction}\label{Radiation Reaction}

\subsection{Equivalence of the Methods}\label{Equivalence of the Methods}

The self-force on a body can be split into two pieces, a
conservative force and a dissipative force, by writing the
retarded field of the source as a sum of parts that are even and
odd under the interchange, {\sl
advanced~$\leftrightarrow$~retarded}, of ingoing and outgoing
radiation.
\begin{equation}
\Phi_{\rm ret} = \frac12(\Phi_{\rm ret}+\Phi_{\rm adv}) +
\frac12(\Phi_{\rm ret}-\Phi_{\rm adv})
\end{equation}
(Because advanced and retarded fields are well-defined on generic,
globally hyperbolic spacetimes, this division is generic for the
self-force arising from linear fields.)  The self-force that we
compute is linear in the field $\Phi$ and can similarly be written
as a sum of a part invariant under the interchange
advanced~$\leftrightarrow$~retarded -- the conservative part of
the force, and a part that changes sign under the interchange --
the dissipative part of the force.

For a charge sustained in perpetual circular orbit by an external
force, the field energy is unchanged from one hypersurface to the
next, and the work done by the self-force is therefore equal to
the energy radiated to the horizon and to infinity. Because the
sign of the radiated energy changes under the
advanced~$\leftrightarrow$~retarded interchange, so does the sign
of the work done by the self-force. Thus the work done by the
self-force is the work done by its dissipative part.

Formally, the stress tensor of a finite test-mass $\mu$ is the sum
of contributions from the mass, the scalar field, and the external
force,
\[
T^{\alpha\beta} = T_\mu^{\alpha\beta} +  T_{\rm S}^{\alpha\beta} +
T_{\rm ext}^{\alpha\beta},
\]
with \be T_{\rm S}^{\alpha\beta} =
\frac1{4\pi}\left(\nabla_\alpha\Phi \nabla_\beta\Phi
    - \frac12 g_{\alpha\beta}\nabla_\gamma\Phi \nabla^\gamma\Phi\right).
\ee Associated with the timelike Killing vector $t^\alpha$ is the
conserved current $j^\alpha=-T^\alpha_{\ \beta}  t^\beta$. The
self-force is constructed from $-\nabla_\beta  T_{\rm
S}^{\alpha\beta} = \rho\nabla^\alpha\Phi$, and the work done by
the self-force between two $t =$ constant hypersurfaces $\Sigma_1$
and $\Sigma_2$ is
\be
  W= \int_\Omega d^4V  \rho\, t^\alpha\nabla_\alpha\Phi
  = \int_\Omega d^4V  \nabla_\alpha j_{\rm S}^\alpha,
\ee
with $\Omega$ the region of spacetime between $\Sigma_1$ and $\Sigma_2$.
We can write this integral as the sum of a vanishing time
derivative and the integral of a 3-dimensional divergence by
writing
\begin{equation}
\nabla_\alpha j_{\rm S}^\alpha = \partial_t j^t + \frac1N
D_a(Nj^a),
\end{equation}
where $N$ is the lapse, $D_a$ the covariant derivative operator on
the hypersurface, and $j_{\rm S}^a$ the projection of $j_{\rm
S}^\alpha$ into the hypersurface. Using
\[
\int_{\Omega} \partial_t j^t d^4V =
\left(\int_{\Sigma_2}-\int_{\Sigma_1}\right)j_{\rm S}^t N d^3V =
0,
\]
we have \be W = \int_{S_\infty} j_{\rm S}^a dS_a + \int_{H} j_{\rm
S}^a N dS_a, \label{flux}\ee where, by $ \int_{S_\infty}$ is meant
the limit $\lim_{r\rightarrow\infty} \int_{S_r}$, with $S_r$ a
sphere of constant coordinate $r$; and $H$ is the horizon.

Because $t^\alpha$ is orthogonal to the horizon and to $S_r$, $W$
has the form
\[
W = \frac1{4\pi}\left[\int_{S_\infty} D^a\Phi\partial_t\Phi dS_a +
\int_{H} D^a\Phi\partial_t\Phi N dS_a\right].
\]
The change of sign of these flux integrals under the
advanced~$\leftrightarrow$~retarded interchange, clear on general
grounds, is apparent from this last form, together with the fact
that the interchange is equivalent in the Kerr geometry to the
transformation induced by the diffeo $t,\phi \leftrightarrow -t,-\phi$.
The explicit form of these integrals for each mode is
given in the next section.

The self-force measured is balanced here by an external force.
When no external force is present, and when one can neglect the change in
the radiative part of the field energy on successive hypersurfaces,
the work $W$ is equal to the change in the
energy of the particle between the hypersurfaces $\Sigma_1$
and $\Sigma_2$.  Modeling the particle by a family of dust
balls with stress-energy
$\rho u^\alpha u^\beta$, with $u^{\alpha}
u^t(t^\alpha+\Omega\phi^\alpha)$, we have
\bea W
&=& \int_\Omega \nabla_\alpha T^\alpha_{\mu\,\beta} t^\beta d^4V
  = \int_\Omega \nabla_\alpha (\rho u^\alpha u_\beta t^\beta) d^4V
\nonumber\\
  &=& \int_\Omega u^\alpha \nabla_\alpha(\rho  u_\beta t^\beta) d^4V.
\label{wdust}\eea

If, along a circular geodesic in the mass, we
define a momentum for which \be p_\alpha t^\alpha = \int_\Sigma
T_{\mu\,\beta}^\alpha t^\beta dS_\alpha
        = \int_\Sigma \rho  u_\beta t^\beta u^t N dV,
\ee then, for a small mass, the integral (\ref{wdust}) is
approximated by \dis W = \int \frac d{d\tau}(p_\beta t^\beta) N
dt, $ and we have \be \frac {dW}{dt} = \frac d{d\tau} p_\alpha
t^\alpha. \ee

Because the dissipative field, $\Phi_{\rm diss} :=
\frac12(\Phi_{\rm ret}-\Phi_{\rm adv})$, is regular for a point
particle, the dissipative part of the self-force is well-defined
for a point particle, without renormalization. The work done by
the self-force at the particle, must then, in our case of
perpetual circular motion, be identical to the radiative flux of
Eq.(\ref{flux}).


\subsection{Energy and Angular Momentum Flux}
We have just argued that we can indirectly find the self-force on
the charge by computing the flux integral of Eq.(\ref{flux}). The
flux of angular momentum to infinity and to the black hole is
similarly given by the integral \be \int_{S_\infty} \widetilde
j_{\rm S}^a dS_a + \int_{H} \widetilde j_{\rm S}^a N dS_a,
\label{Jflux}\ee with $\widetilde j_{\rm S}^a = T^\alpha_\beta
\phi^\beta$, with $\phi^\alpha$ the rotational Killing vector
$\partial_\phi$.

Again, when no external torque is present and when the difference
between the field angular momentum on successive hypersurfaces can
be neglected, the radiated angular momentum is equal to the change
in the angular momentum of the particle.

We denote by $E=-u_t=u_\alpha t^\alpha$ and $L=u_\phi=u_\alpha
\phi^\alpha$ the energy and (z component of) angular momentum per
unit rest mass of the charge. In the full dynamical problem, no
flux reaches spatial infinity, and one relates a change in energy
between successive hypersurfaces to asymptotic flux by choosing a
family of asymptotically null hypersurfaces.  In our model,
however, the charge has been orbiting forever, and the integrals
can be evaluated at spatial infinity and at the bifurcation
horizon (as in the previous section). Consider first the flux of
angular momentum $L$.  We compute Eq.(\ref{Jflux}) to get ${dL
\over dt}$ (times the rest mass $\mu$), for the particle. For a
surface of constant $r$ (and $t$), the flux in the direction of
increasing $r$ is
\begin{equation}\label{Flux through surface of constant r}
\mathcal{L} = \int {T^r _{\mbox{ }\phi} } \Sigma d\Omega = \int
{T_{r\phi}} \Delta d\Omega.
\end{equation}

The real scalar field is a sum $\Phi = \sum\limits_{lm} {\Phi_{lm}
}$ of complex fields whose angular eigenfunctions
$S_{lm}e^{im\phi}$ are orthogonal.  The flux integral is then sum
of integrals for each mode, and each mode integral involves a
stress-energy tensor of the form
\begin{equation}\label{Stress-energy Tensor}
T_{\alpha \beta } =\frac1{4
\pi}\left(\nabla_{(\alpha}\Phi_{lm}^* \nabla_{\beta)}\Phi_{lm}
 - \frac12 g_{\alpha\beta}\nabla_\gamma\Phi_{lm}^* \nabla^{\gamma}\Phi_{lm}
 \right).
\end{equation}

The modes have asymptotic behavior
\begin{align}\label{Modes End Behavior--horizon}
\Phi _{lm} &= \frac{q}{{2\pi u^t }}C_{lm}^ +  S_{lm} \left(
{{\textstyle{\pi  \over 2}}} \right)S_{lm} \left( \theta
\right)e^{-im(\Omega  - \omega _ +  )r^*}e^{im(\phi  - \Omega t)}
+O(r - r_ + )\qquad &(\mbox{near } r_+)&\\
\label{Modes End Behavior--infinity} \Phi _{lm} &= \frac{q}{{2\pi
u^t }}C_{lm}^\infty S_{lm} \left( {{\textstyle{\pi  \over 2}}}
\right)S_{lm} \left( \theta \right)\frac{1}{r}e^{im\Omega
r^*}e^{im(\phi  - \Omega t)} +O(r^{-2})\qquad
    &(\mbox{near } \infty)&.
\end{align}
The ($r-\phi$) component of the stress energy in each mode has
corresponding behavior
\begin{align}\label{Stress-energy mode horizon}
\left( {T_{r\phi } } \right)_{lm}
 &= \frac{1}{4 \pi} \frac{{r^2 + a^2 }}{\Delta } \left( {\Omega  -
\omega _ +  } \right) m^2 \Phi _{lm}\left[1+O(r - r_ +
)\right]\qquad &(\mbox{near } r_+)&\\
\label{Stress-energy mode infinity}
\left( {T_{r\phi } } \right)_{lm}
 &= \frac{-1}{4 \pi} \frac{{r^2  + a^2 }}{\Delta } \Omega
m^2 \Phi _{lm} \left[1+O(r^{-1})\right]\qquad
    &(\mbox{near } \infty)&,
\end{align}
and by (\ref{Flux through surface of constant r}) the fluxes are

\begin{align}\label{Flux horizon}
\mathcal{L}(r_+) &= \frac{1}{2} \left( \frac{q}{{2\pi u^t
}}\right)^2 \left( {\Omega  - \omega _ + } \right)(2Mr_ +) m^2
\left| {C_{lm}^ +  } \right|^2 S_{lm} \left( {{\textstyle{\pi
\over 2}}} \right)^2\\
\label{Flux infinity} \mathcal{L}(\infty) &= - \frac{1}{2} \left(
\frac{q}{{2\pi u^t }}\right)^2 \Omega m^2 \left| {C_{lm}^ \infty }
\right|^2 S_{lm} \left( {{\textstyle{\pi \over 2}}} \right)^2.
\end{align}

These give the rate at which angular momentum crosses the surfaces
in the direction of increasing $r$.  The flux integrals
(\ref{flux}), (\ref{Jflux}) involve the outward normal of a
3-volume bounded by spatial infinity and the horizon, implying a
sign change for the contribution (\ref{Flux horizon}) at the
horizon:
\begin{equation}\label{dLdt1}
\frac{{dL}}{{dt }} = \frac{-q^2}{8 \pi^2 \mu \left(u^t\right)^2}
\sum\limits_{l,m} {m^2 S_{lm} \left(
{\textstyle\frac\pi2}\right)^2 \left[ {2Mr_ + \left( {\Omega  -
\omega _ +  } \right)\left| {C_{lm}^ + } \right|^2  + \Omega
\left| {C_{lm}^\infty  } \right|^2 } \right]}.
\end{equation}

For $E$, a similar calculation yields
\begin{equation}\label{dEdL}
\frac{{dE}}{{dt }}= \Omega \frac{{dL}}{{dt }}.
\end{equation}

This relationship is in fact necessary for the particle to remain
in circular orbit as it radiates and spirals inward.

\subsection{Half-Retarded Minus Half-Advanced}

Here we calculate the dissipative self-force directly, by taking
the gradient of the half-retarded-minus-half-advanced solution.
The expression of (\ref{Field Sum}) is the retarded solution. As
noted above, the advanced solution is obtained from the retarded
by the transformation $\phi \to -\phi$, $t \to -t$.  Denoting the
retarded solution by $(-)$ and the advanced solution by $(+)$, we
have
\begin{equation}\label{Field Sum retadv}
\Phi_\pm  = \frac{q}{{2 \pi u^t }}\sum\limits_{l,m} {R_{lm} }
(r)S_{lm} ({\raise0.5ex\hbox{$\scriptstyle \pi $}
\kern-0.1em/\kern-0.15em \lower0.25ex\hbox{$\scriptstyle
2$}})S_{lm} (\theta )e^{\mp im(\phi  - \Omega t)},
\end{equation}
and the dissipative solution is then
\begin{equation}\label{Field Sum dis}
\Phi_{\textrm{dis}} = {1 \over 2}\left(\Phi_--\Phi_+\right) =
\frac{q}{{2 \pi u^t }}\sum\limits_{l,m} i{R_{lm} } (r)S_{lm}
({\raise0.5ex\hbox{$\scriptstyle \pi $} \kern-0.1em/\kern-0.15em
\lower0.25ex\hbox{$\scriptstyle 2$}})S_{lm} (\theta )\sin \left[
m\left(\phi-\Omega t\right)\right].
\end{equation}

To elucidate this expression, note the $m\to-m$ symmetry
properties of its components.  Eqs. (\ref{Modes End
Behavior--horizon}) and (\ref{Modes End Behavior--infinity}) show
that $R_{l\left(-m\right)}=R_{lm}^*$.  The $S_{lm}$ were chosen to
have the same sign as the Legendre functions $P^m_l$, so
$S_{l\left(-m\right)}=\left(-1\right)^m S_{lm}$.  The sine
function of course changes sign with m.  This all implies that the
imaginary part of each mode of $\Phi _{{\rm{dis}}}$ (coming from
the \textit{real} part of $R_{lm}$) changes sign with $m$ while
the real part doesn't.  Since $m=0$ does not contribute to the sum
by virtue of the sine, the solution is real (like the retarded and
advanced solutions), and more importantly comes completely from
the imaginary parts of the radial functions, which, as solutions
to the \textit{homogeneous} radial equation, are smooth.  We thus
obtain the more transparent version of $\Phi_{\textrm{dis}}$,
\begin{equation}\label{Field Sum dis2}
\Phi_{\textrm{dis}} = \frac{-q}{{2 \pi u^t }}\sum\limits_{l,m}
\textrm{Im}\left[{R_{lm}} (r)\right] S_{lm}
({\raise0.5ex\hbox{$\scriptstyle \pi $} \kern-0.1em/\kern-0.15em
\lower0.25ex\hbox{$\scriptstyle 2$}})S_{lm} (\theta )\sin \left[
m\left(\phi-\Omega t\right)\right],
\end{equation}
each term of which is smooth at the particle. The force is given
by minus the orthogonal projection of the gradient of this
quantity (evaluated at the location of the particle), times the
scalar charge $q$.  To wit,

\begin{equation}\label{Force}
\frac{d}{{d\tau }}\left( {p_\alpha  } \right) = -q(\delta _\alpha
^\beta   + u_\alpha  u^\beta  )\nabla _\beta  \Phi _{{\rm{dis}}}.
\end{equation}

The location of the particle is $\phi = \Omega t$, so the $r$ and
$\theta$ components vanish by virtue of the sine in Eq.(\ref{Field
Sum dis}). The $t$ and $\phi$ components give us the evolution of
the conserved quantities $E$ and $L$. Because $\nabla_t=-\Omega
\nabla_\phi$, the projection operator is just the identity (ie,
$(\delta _\alpha ^\beta   + u_\alpha u^\beta )\nabla
_\beta=\nabla_\alpha$), and Eq.(\ref{Force}) yields

\begin{equation}\label{dLdt2}
\frac{{dL}}{{dt }} = \frac{{ - q^2 }}{{2\pi \mu \left(u^t\right)^2
}}\sum\limits_{l,m} {m{\mathop{\rm Im}\nolimits} \left[ {R_{lm}
\left( {r_0 } \right)} \right]} S_{lm} \left(
{{\raise0.5ex\hbox{$\scriptstyle \pi $} \kern-0.1em/\kern-0.15em
\lower0.25ex\hbox{$\scriptstyle 2$}}} \right)^2
\end{equation}
\begin{equation}\label{dEdL2}
(\frac{{dE}}{{dt }}= \Omega \frac{{dL}}{{dt }}).
\end{equation}

We have used $\frac{{d}}{{d\tau}}=u^t\frac{{d}}{{dt}}$ to rewrite
for easy comparison with Eq.(\ref{dLdt1}). Eqs. (\ref{dLdt1}) and
(\ref{dLdt2}) must be equal, and furthermore the equality must be
mode by mode, since the argument of section \ref{Equivalence of
the Methods} is valid mode by mode. We can therefore write

\begin{equation}\label{mode by mode}
 4 \pi {\mathop{\rm Im}\nolimits}\left[ {R_{lm} \left( {r_0 } \right)}
\right] = m\left[ {2Mr_ +  (\Omega  - \omega _ +  )\left| {C_{lm}^
+  } \right|^2  + \Omega \left| {C_{lm}^\infty  } \right|^2}
\right],
\end{equation}
and in fact this relationship provides a convenient means to
estimate the total numerical error involved in the computation of
a given $R_{lm}$.

\section{Discussion}\label{Discussion}

\subsection{The Radial Functions}

\subsubsection{Computation}

The general scheme for the computation of the radial functions is
presented in Section \ref{Integration of the Radial Equation}. To
summarize, one integrates in from infinity and out from the
horizon, using asymptotic series to provide initial values, and
then patches the two solutions together at the particle.  Here we
discuss our actual implementation of this method.  There are two
sources of error in the computation: the integration itself, and
the initial value that starts it (gotten from the series).  We
pick a desired maximum error for the whole computation and make
sure that the error from neither source exceeds this bound.  For
the integration, we use the Runga-Kutta Cash-Karp order 4/5
routine, as implemented in the GSL scientific library \cite{gsl},
which reports the error introduced at each step (note that we
integrate in $r$, not $r*$). The sum of the magnitudes of these
errors is a (gross) upper bound on the total numerical error, and
we adjust the parameters of the integrator until this falls below
our desired maximum. We control the error introduced by the
initial value by increasing the value's accuracy until the result
of the integration (at $r_0$) does not change to within the
desired error. One can increase the accuracy either by increasing
the number of terms in the series used or by moving further away
from the particle, where the series are increasingly convergent.
We chose to fix the number of terms and increase the distance. For
the results accurate to six significant figures presented in this
paper, we used fifteen terms from each series, allowing us to
begin most infinity integrations somewhere between $r/M=100$ and
$r/M=1000$, and most horizon integrations at distances of order
$10^{-2}$ to $10^{-3}$ away from the horizon. The series appear to
converge slowest (i.e., we must begin furthest from the particle)
when the difference between $l$ and $|m|$ is largest.  The
patching of the solutions requires the calculation of the
Wronskian; although this can be done at the single point $r_0$, we
integrate a little past $r_0$ in either direction to have a range
of values over which we can verify that the quantity $\Delta
W(R^+_{lm},R^\infty_{lm})$ of Eq.(\ref{Patch}) is constant (to the
desired accuracy). Finally, at the end of each integration we
check that Eq.(\ref{mode by mode}) is satisfied (again to the
desired accuracy).

\subsubsection{ Properties }

Although the field they sum to is real, the radial functions
themselves are complex-valued.  At the particle, the real parts
have cusps while the imaginary parts remain smooth (see Fig.
\ref{fig:Radial Functions} for a picture). This is because the
delta function in our source is purely real.  The imaginary parts
can be thought of as solutions to the source-free wave equation,
coupled to our problem via the radiative boundary conditions. Only
these smooth functions contribute to the dissipative field (see
Eq. \ref{Field Sum dis2}). Away from the particle, the radial
functions can be understood in terms of the asymptotic series Eqs.
(\ref{Radial Series Horizon}) and (\ref{Radial Series Infinity}).
The first terms of these series describe the end behavior,

\begin{align}\label{Horizon End Behavior}
R_{lm}\left(r \to r_+\right) &\sim e^{ - im(\Omega  - \omega _ +
)r*},\\
\label{Infinity End Behavior}  R_{lm}\left(r \to \infty \right)
&\sim {1 \over r} e^{ im\Omega r^*}.
\end{align}

To understand these we must `translate' from $r*$ to $r$. As $r
\to \infty$, $r* \to r$, so the behavior at infinity is simply a
$1/r$ decaying sinusoid of frequency $m \Omega$.  As $r \to r_+$,
however, $r* \to -\infty$. This means that the distance between
points in $r$ is increasingly stretched by $r*$ as one approaches
the horizon. The distance between wave crests is constant in $r*$,
so in $r$ it is steadily decreasing. The effect is that wavefronts
`pile up' on the horizon, as in Fig. \ref{subfig:horizon} below.

\begin{figure}[h]
\subfigure[$R_{11}$]{\label{subfig:R11}
\includegraphics[scale=.6]{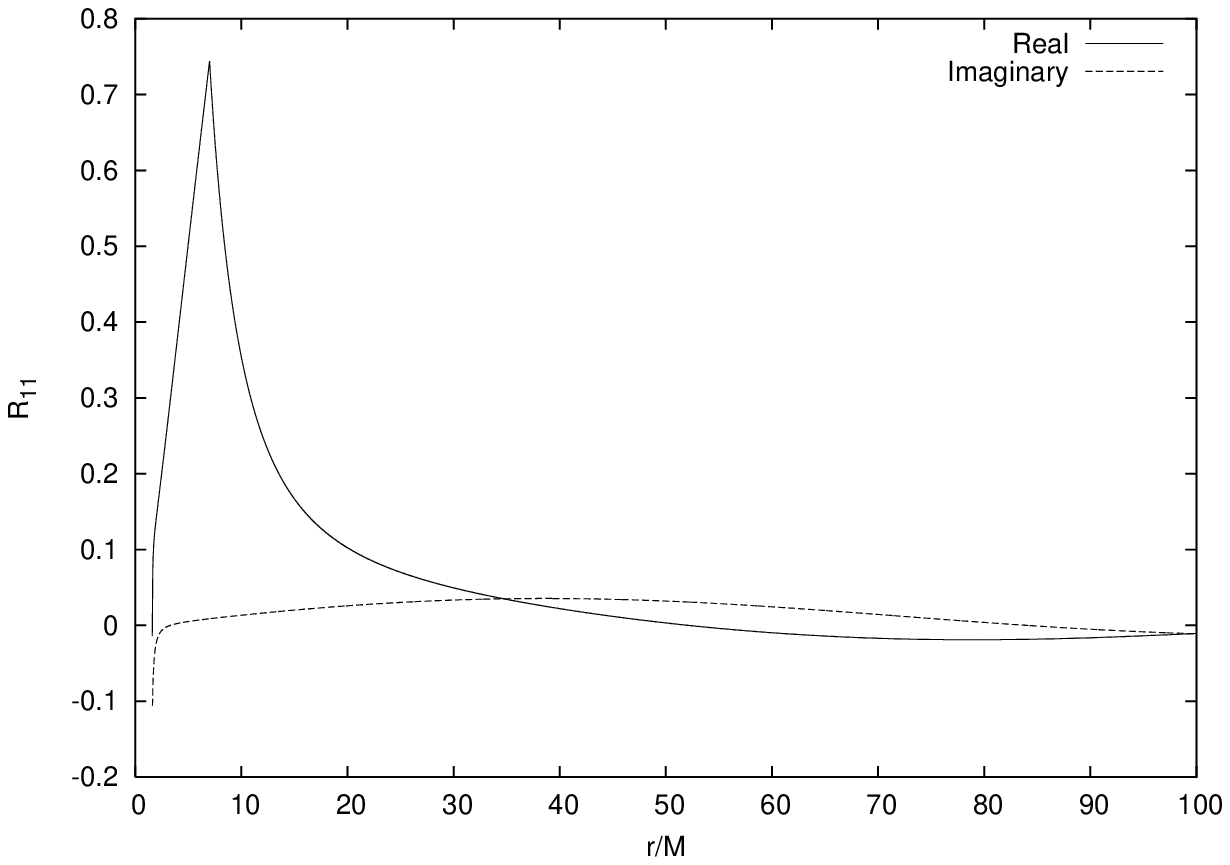}}
\subfigure[$R_{10,10}$]{\label{subfig:R1010}
\includegraphics[scale=.6]{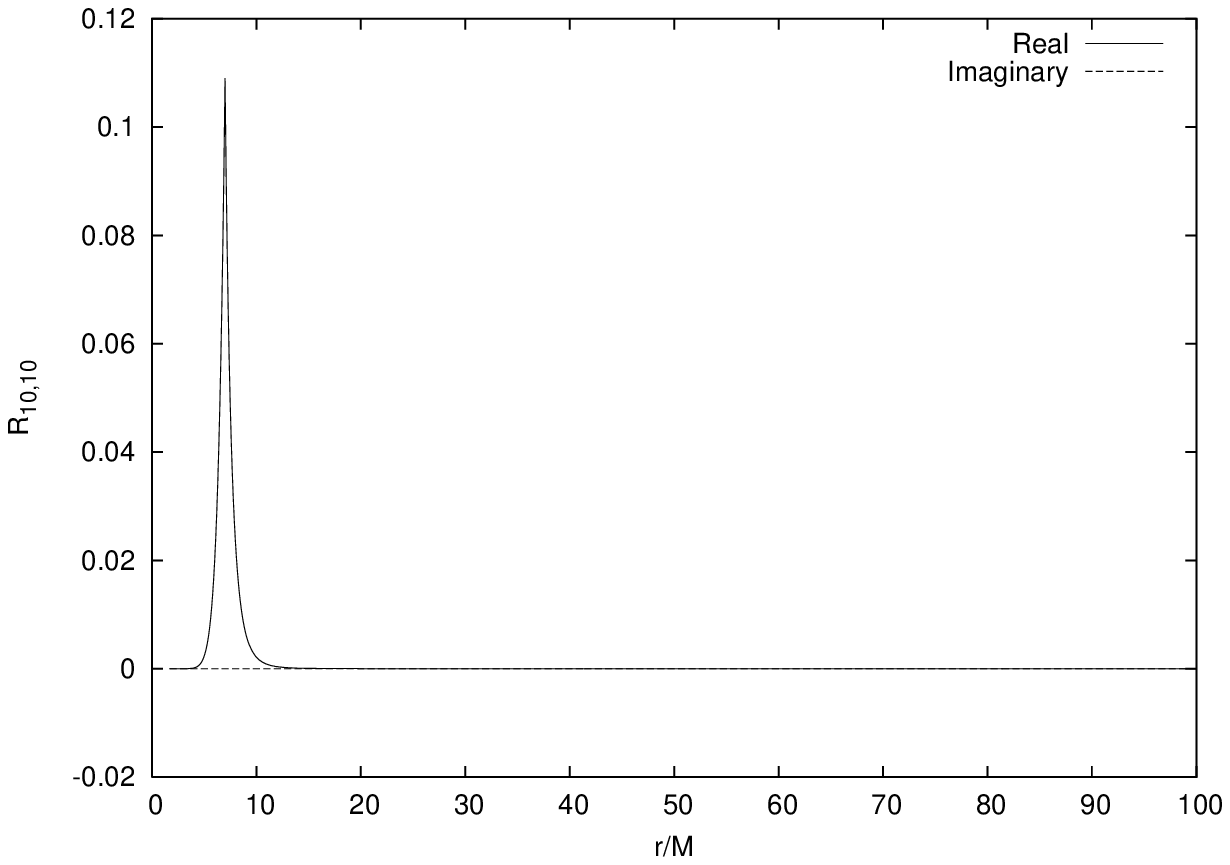}}
\caption{Two Radial Functions, for $a=.8M$, $r_0=7M$ (prograde
orbit, $\Omega \approx .05$).  The real parts have cusps at the
position of the particle, while the imaginary parts are smooth
everywhere.  The oscillations in (b) are too small compared to the
cusp to be seen (we show these below).} \label{fig:Radial
Functions}
\end{figure}

\begin{figure}[h]
\subfigure[Near the horizon,
$r_+=1.6M$]{\label{subfig:horizon}\includegraphics[scale=.6]{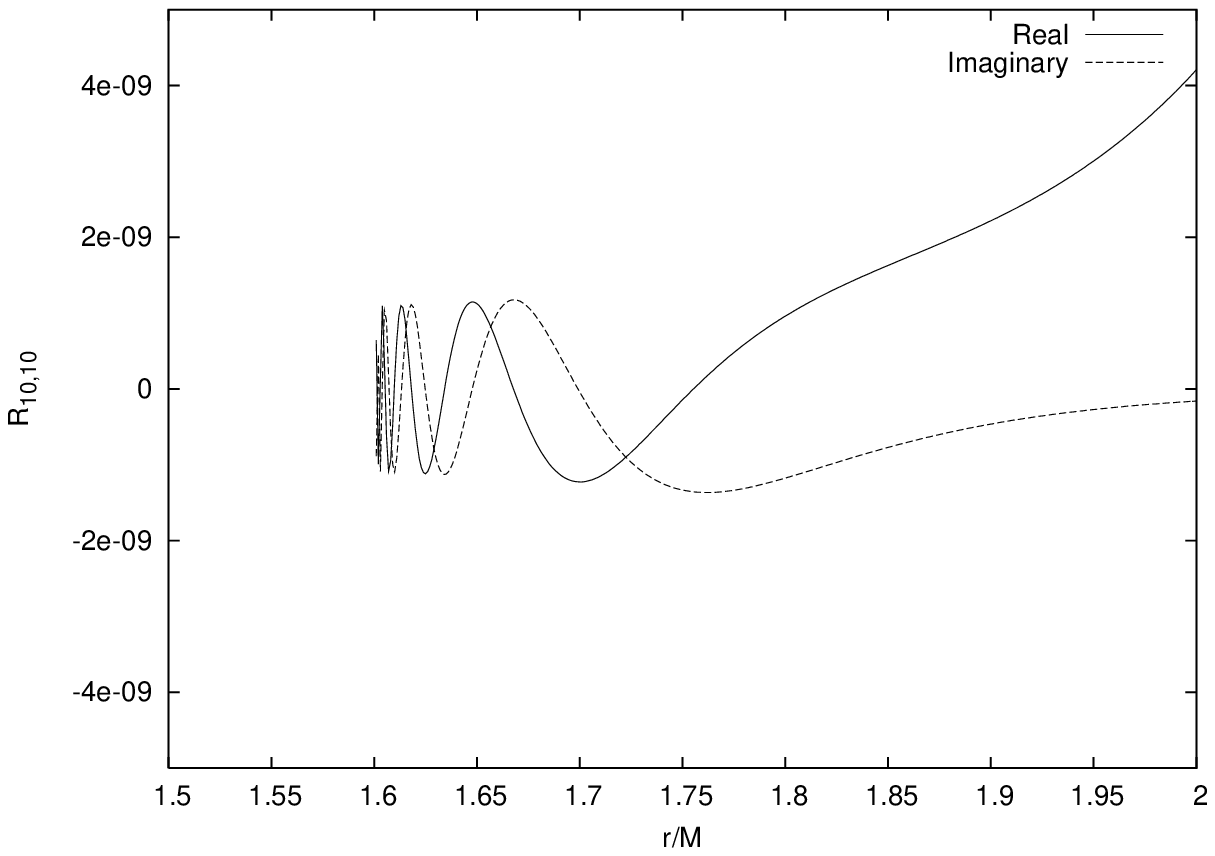}}
\subfigure[Seeing the infinity behavior]{\label{subfig:infinity}
\includegraphics[scale=.6]{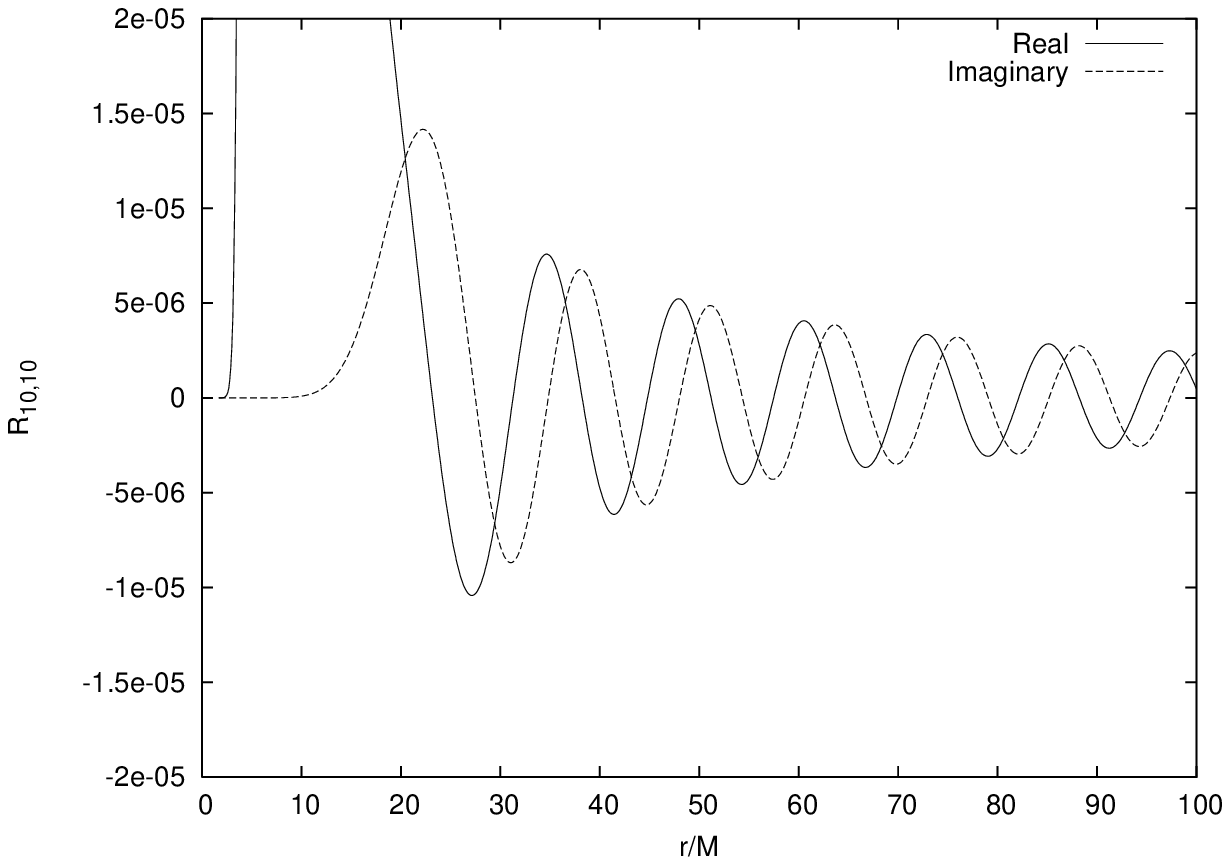}}
\caption{Closer looks at $R_{10,10}$ of fig. \ref{subfig:R1010}.
At the horizon wave fronts pile up, and at infinity we have a
decaying sinusoid.} \label{fig:R1010 Closeups}
\end{figure}

\subsection{Dissipative Self-force}

The dissipative self-force is characterized by $dL \over dt$,
calculated from Eq.(\ref{dLdt1}) or equivalently Eq.(\ref{dLdt2}).
A particle in the adiabatic regime will inspiral along circular
orbits according to this quantity until it begins a transition to
a `plunge' orbit near the inner-most stable circular orbit (ISCO),
as discussed in \cite{oriThorne2000}.  Thus it is only meaningful
to compute results for orbits outside the ISCO.  The convergence
of the $l,m$ sums that give $dL \over dt$ is quite good, until the
particle gets too close to the black hole.  For most cases, the
ISCO is far enough out that all orbits of interest require a small
number ($l=10-15$) of modes to converge to six significant
figures. However, as a co-rotating hole moves the ISCO inwards
(see \cite{bardeenPressTeukolsky1972} for a nice discussion of
circular orbits in Kerr), more modes can be required.  For the
extremal co-rotating case, the ISCO approaches the horizon at
$r=M$ and the adiabatic regime involves orbits whose convergence
is quite bad. For the $r=2M$ orbit result shown in fig.
\ref{fig:dLdt data} below, $l=30$ modes were required for the six
significant figures.

\begin{figure}[h]
\includegraphics[scale=.6]{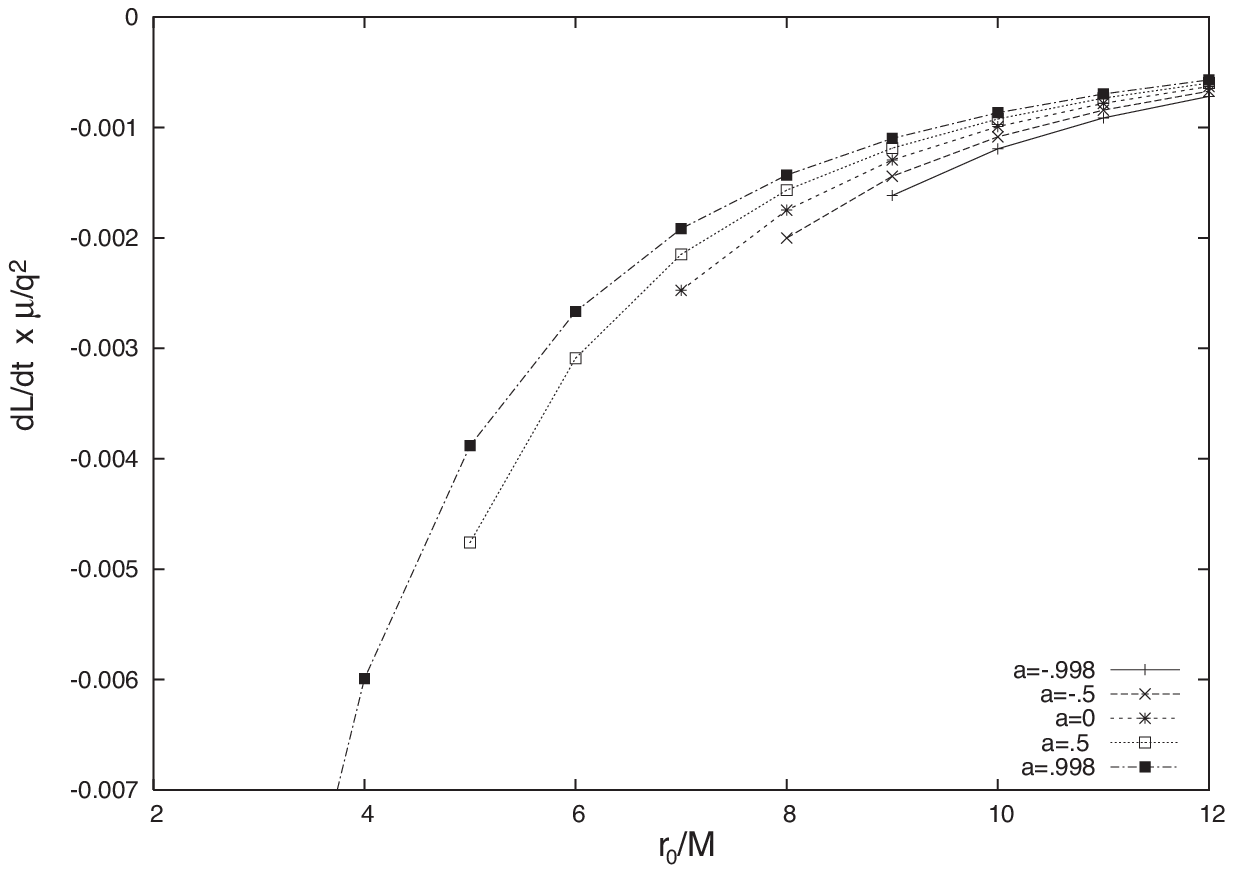}
\begin{tiny}
\begin{tabular}{|c|c|c|c|c|c|c|c|c|c|c|c|}
\hline
$a$/$r_0$ & 2 & 3 & 4 & 5 & 6 & 7 & 8 & 9 & 10 & 11 & 12 \\
\hline -.998 & & & & & & & & -1.61878e-03 &  -1.19081e-03 &
-9.10241e-04 &
-7.16358e-04 \\
\hline -.5 & & & & & & & -1.99659e-03 & -1.43831e-03 &
-1.08177e-03 &
-8.40224e-04 & -6.69240e-04 \\
\hline 0 & & & & & & -2.47425e-03 & -1.74808e-03 &  -1.29433e-03 &
-9.92217e-04
& -7.81459e-04 & -6.29011e-04 \\
\hline .5 & & & & 4.75318e-03 & -3.08374e-03 & -2.14376e-03 &
-1.56388e-03 & -1.18299e-03 & -9.20761e-04 & -7.33412e-04 & -5.95476e-04 \\
\hline .998 & -1.68271e-02 & -9.89666e-03 & -5.98924e-03 &
-3.88003e-03 & -2.66363e-03 & -1.91487e-03 & -1.42814e-03 &
-1.09729e-03 & -8.63947e-04 & -6.94251e-04 & -5.67603e-04 \\
\hline
\end{tabular}
\end{tiny}
\caption{Some values of ${dL \over dt} \times {\mu \over q^2}$ for
various orbits. Negative values of $a$ represent retrograde
orbits. Results for orbits within the inner-most stable circular
orbit are unphysical and not shown.  All results are accurate to
the six significant figures displayed.} \label{fig:dLdt data}
\end{figure}

\subsection{Acknowledgements}
We thank Eirini Messaritaki for helpful discussions.  SG thanks
Vincent Moncrief for sponsoring this work as a senior thesis at
Yale, and for assistance with conceptual and technical
difficulties during the course of the work. This work was
supported in part by NSF grants 0243631, 0071044, and 0200852.

\section{Appendix--The Recursion Relations}\label{sec:Recursion
Relations} The $C_n^ +$ and $C_n^ \infty$ of Eqs. (\ref{Radial
Series Horizon}) and (\ref{Radial Series Infinity}) satisfy
recursion relations, which were derived with the help of
\textit{Mathematica}.  The procedure is to plug in the forms of
Eqs. (\ref{Radial Series Horizon}) and (\ref{Radial Series
Infinity}) to the radial equation (\ref{Radial Equation}), and
rewrite the results with a single summation index, so that one
derives an expression that must be zero (the equation is
homogeneous away from the particle). These expressions are
\begin{eqnarray}\label{Recursion Relations}
\sum\limits_{i = 0}^4 {f_i^ +  C_{n - i}^ +  } &=& 0 \\
\sum\limits_{i = 0}^5 {f_i^ \infty  C_{n - i}^ \infty  } &=& 0,
\end{eqnarray}

with $f_i^+$ and $f_i^\infty$ given below (where we have defined
$\Omega_+\equiv\Omega-\omega_+)$.

\begin{eqnarray*}
f_0^+ &=& -4am^2M\,r_+\Omega  + 4M^2 \left[ n^2 -
2imnr_+\,\Omega_+ + 2m^2Mr_+\left( {\Omega }^2 - \Omega_+^2
\right) \right] \\
& & + a^2\left[ -4n^2 + 8imMn\Omega_+ + m^2\left( 1 - 4M^2\left(
{\Omega
}^2 - \Omega_+^2 \right)  \right)  \right] \\
f_1^+ &=& -2r_+\left( -1 + 3n - 2n^2 + \lambda  + a^2m^2{\Omega
}^2 \right)  + 4i a^2m\left( -1 + 2n \right) \Omega_+ -
16m^2M^2r_+\left( {\Omega }^2 - {\Omega_+}^2 \right) \\
& & + 2M\left[ 1 + 2n^2 - \lambda  + 2am^2\Omega  +
3a^2m^2{\Omega}^2 - 4i mr_+\Omega_+ - 4a^2m^2{\Omega_+}^2 +
n\left( -3 + 6i mr_+\Omega_+
\right)  \right] \\
f_2^+ &=& 2 + n^2 - \lambda  - 5a^2m^2{\Omega }^2 +
12m^2Mr_+{\Omega }^2 - 4i mM\Omega_+ + 10i mr_+\Omega_+ +
4a^2m^2{\Omega_+}^2 \\
& & - 12m^2Mr_+{\Omega_+}^2 + i n\left[ 3i  + 4m\left( M - 2r_+
\right)
\Omega_+ \right] \\
f_3^+ &=& 2m\left[ -i \left( -2 + n \right) \Omega_+ + 2mr_+\left(
{\Omega }^2 - {\Omega_+}^2 \right)  \right] \\
f_4^+ &=& m^2\left( {\Omega }^2 - {\Omega_+}^2 \right) \\
\\
f_0^\infty &=& -2i m\left( -1 + n \right) \Omega \\
f_1^\infty &=& 2 + n^2 - \lambda  - 8i mM\Omega  - a^2m^2{\Omega
}^2 +
n\left( -3 + 4i mM\Omega  \right) \\
f_2^\infty &=& 2\left[ i a^2m\left( 5 - 2n \right) \Omega  +
M\left( -10 + 9n - 2n^2 + \lambda  - 2am^2\Omega  + a^2m^2{\Omega
}^2 \right)
\right] \\
f_3^\infty &=& 4M^2{\left( -3 + n \right) }^2 - a^4m^2{\Omega }^2
+ a^2\left[ 18 + m^2 - 12n + 2n^2 - \lambda  + 4i mM\left( -3 + n
\right)
\Omega  \right] \\
f_4^\infty &=& 2a^2\left( -4 + n \right) \left[ M\left( 7 - 2n
\right)
 - i a^2m\Omega  \right] \\
f_5^\infty &=& a^4\left( -5 + n \right) \left( -4 + n \right) \\
\end{eqnarray*}


\begin{thebibliography}{99}

\bibitem{ligo} The LIGO Scientific Collaboration: B. Abbott, et al,
Nucl.Instrum.Meth. {\bf A517} (2004) 154-179, and reference
therein.

\bibitem{geo} B.~Wilke, et al.,
Class. Quant. Grav. {\bf 9(7)} (2002) 1377.

\bibitem{tama} http://tamago.mtk.nao.ac.jp/.

\bibitem{lisa} See http://lisa.jpl.nasa.gov/

\bibitem{biww} L.~Blanchet, B.~Iyer, C.M.~Will and A.G.~Wiseman,
Class.Quant.Grav. {\bf 13} (1996) 575-584.

\bibitem{bcv} A.~Buonanno, Y.~Chen, Y.~Pan, M.~Vallisneri,
Phys.Rev. {\bf D70} (2004) 104003.

\bibitem{Dirac38} P. A. M. Dirac, Proc. Roy. Soc. \textbf{A167}, 148
(1938).

\bibitem{DB60} B. S. DeWitt and R. W. Brehme, Ann. Phys. (N.Y.)
\textbf{9},
   220 (1960).

\bibitem{Hobbs68} J. M. Hobbs, Ann. Phys. \textbf{47}, 141 (1968).

\bibitem{MST97} Y. Mino, M. Sasaki and T. Tanaka, Phys.\ Rev.\ D \textbf{55},
   3457 (1997).

\bibitem{QW97} T. C. Quinn and R. M. Wald, Phys.\ Rev.\ D \textbf{56} 3381,
   (1997).

\bibitem{Quinn00} T.C. Quinn, Phys.Rev. D \textbf{62}, 064029 (2000).

\bibitem{BO01} L.~Barack, A.~Ori, Phys.Rev. {\bf D64} (2001) 124003.

\bibitem{hughes01} S.A.~Hughes,
Phys.Rev. {\bf D61} (2000) 084004; Erratum-ibid. D63 (2001) 049902

\bibitem{poissonSasaki95} E.~Poisson and M.~Sasaki,
Phys.Rev. {\bf D51} (1995) 5753-5767

\bibitem{wiseman00} A.G.Wiseman, Phys. Rev. D \textbf{61} (2000) 084014.

\bibitem{mtw}
C.W. Misner, K.S. Thorne and J.A. Wheeler, {\it Gravitation}
(Freeman, San Francisco, 1973).

\bibitem{numrec} W.H. Press, S.A. Teukolsky, W.T. Vettering
and B.P. Flannery, {\it Numerical Recipes}, (Cambridge University
Press, 1992).


\bibitem{carter68} B. Carter, Phys. Rev. {\bf 174}, 1559 (1968).

\bibitem{TeukolskyPaper}
S. Teukolsky, Phys. Rev. Lett. {\bf 29}, 1114 (1972).

\bibitem{gsl}
http://www.gnu.org/software/gsl/

\bibitem{oriThorne2000}
Amos Ori, and Kip S. Thorne, Phys.Rev. {\bf D62} (2000) 124022.

\bibitem{bardeenPressTeukolsky1972}
J.M. Bardeen, W.H. Press and S.A. Teukolsky, Astrophys.J., {\bf
178}:347-369, 1972

\end{thebibliography}
\end{document}